\documentclass{aa}
\usepackage{graphicx}
\usepackage{txfonts}
\begin{document}
   \title{2.3\,$\mu$m CO emission and absorption from \\
       young high-mass stars in M17
   \thanks{Based on observations made
       with ESO telescopes at the Paranal Observatory under programme
       IDs 069.C-0303, 073.C-0170, and 075.C-0381.}}

\titlerunning{CO from Young Stars in M17}
\authorrunning{Hoffmeister et al.}

\author{V.H. Hoffmeister\inst{1}, R. Chini\inst{1}, C.M. Scheyda\inst{1}
    \and
    D. N\"urnberger \inst{2}
    \and
    N. Vogt \inst{3,4}
    \and M. Nielbock
    \inst{1}
    }

   \offprints{{\tt vhoff@astro.rub.de}}

\institute{Astronomisches Institut, Ruhr-Universit\"at Bochum, 44780
    Bochum, Germany
    \and
    European Southern Observatory, Santiago 19, Chile
    \and
    Instituto de Astronom{\'\i}a, Universidad Cat{\'o}lica del Norte,
    Antofagasta, Chile \and Departamento de F{\'\i}sica y Meteorolog{\'\i}a,
    Universidad de Valpara{\'\i}so, Chile}

\date{Received 30 June 2006; accepted 25 July 2006}

  \abstract
   {}
{We are studying the extremely young cluster of M17 to
investigate the birth of high-mass stars and the initial mass
function.}
{Deep $JHKL$ imaging and $K$-band spectroscopy from the VLT of 201
stars toward the cluster is presented.}
{The majority of 104 stars show the CO band-head in absorption. Half
of them emit X-rays and/or have infrared excess, indicative of
very young objects. Their intrinsic IR luminosity is compatible with
intermediate and high-mass pre-main sequence stars. Nine additional
stars have the CO feature in emission, while sixty sources are
lacking any stellar spectral feature due to veiling by circumstellar
dust.}
{We suggest that CO absorption is -- as in the case of low-mass
stars -- also a common feature during the early evolution of stars
with higher masses. According to model calculations the observed CO
absorption is most likely a sign of heavily accreting protostars
with mass accretion rates above $10^{-5}\,$M$_\odot$\,yr$^{-1}$.}


   \maketitle

\section{Introduction}

Numerous young stellar objects (YSOs) display the CO band-heads at
$2.3$ $-$ $2.4\,\mu$m in emission. There is consensus that this CO
emission originates in extremely dense ($n_{\rm H} \ge 10^{10}$
cm$^{-3}$) and warm (1500\,K $<$ $T$ $<$ 4500\,K) regions associated
with YSOs and is likely a result of the disk accretion process
during early stellar evolution. A variety of mechanisms and
models have been proposed to explain their precise origin. These
include circumstellar disks, stellar winds, magnetic accretion
mechanisms, and inner disk instabilities
(e.g., Carr 1989; Carr et al. 1993; Chandler et al. 1993; Biscaya et
al. 1997). Recently, the CO emission has been used to infer
properties of the associated circumstellar disks
(Bik \& Thi 2004; Blum et al. 2004).

CO band-heads in absorption, in contrast, are typical of the
photospheres of cool, mostly evolved stars. Nevertheless, initial
surveys of star--forming regions
(e.g., Casali \& Eiroa 1996) have demonstrated that
low-luminosity Class\,II sources, i.e., stars with flat or decreasing
spectral energy distributions (SEDs), such as T\,Tauri stars, may also
display strong CO absorption features. Sources with steeply rising
SEDs (Class\,I) showed much weaker, or even undetectable, CO
absorption. \cite{greene96} also found that the CO bands are much
weaker in Class\,I sources. On the other hand, CO absorption arising
from an expanding shell or from an accretion disk has been discussed
in the context of FU Orionis objects
(Hartmann et al. 2004).

Models have shown that variations in the strength and the appearance
of the CO feature -- either in emission or in absorption -- depend
on parameters like the stellar temperature, the geometry of the
circumstellar material, and the mass accretion rate onto the
protostellar object
(Carr 1989; Calvet et al. 1991). The weakness or even the absence of
any CO feature may therefore be a complex combination of various
competing mechanisms.

We investigate the stellar content of the extremely young cluster NGC
6618 that is exciting the luminous \ion{H}{ii} region of M17 at a 
distance of 1.9\,kpc (Schmidt et al., in prep.). As we will show in
the following, we have found a considerable population of CO sources
in this cluster. Only a few of them show the CO band-head in
emission (COES); the majority of objects display CO in absorption
(COAS). Associated X-ray emission and IR excess suggest that the
youngest generation of intermediate and high-mass stars is in the
process of heavy accretion.

\section{Observations}

We have performed both near-infrared imaging and spectroscopy of the
M17 cluster. The images have been obtained with ISAAC ($JHK_s$ and
$L_p$, hereafter $JHKL$) at the ESO VLT in September 2002 and September 2004,
respectively. Details of the observation and reduction procedures
will be described elsewhere (Hoffmeister et al., in prep.; Scheyda
et al., in prep.). The $1.00 - 1.04\,\mu$m spectra were obtained
from May to August 2005, and the $2.0 - 2.4\,\mu$m spectra in August
2004 as well as from May to July 2005 with ISAAC; the spectral
resolutions are 5700 and 1500, respectively.

The selection of the sources across the cluster field was fairly
arbitrary. We centred one or two sources onto the $120\,\arcsec$
slit, while on average two to three additional sources happened
to be covered by the same slit position. In this way, we obtained
$K$-band spectra for 201 stars of which two--thirds appear in our
sample by pure chance. As such, this survey can be regarded as
fairly unbiased.

\begin{figure}
\resizebox{\hsize}{!}{\includegraphics*[width=\textwidth]{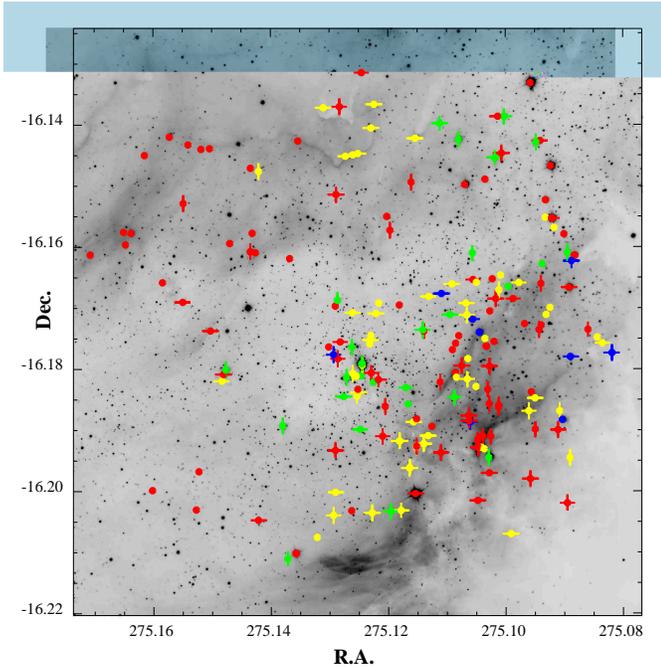}}
\caption{$K$-band mosaic of the cluster region in M17.
All sources with $K$-band spectra are marked. Red: CO absorption,
blue: CO emission, yellow: featureless, green: others. X-ray
emission is indicated by a vertical bar, IRE by a horizontal bar.
\label{f1}}
\end{figure}

\section{Results}
\label{results}

Figure\,\ref{f1} shows a $K$-band image of the cluster field in M17
(FOV $330" \times 342"$) with about 18.000 sources down to a
limiting magnitude of $K = 19.0\,$mag (Hoffmeister et al., in
prep.). The region was also observed in the $L$-band (Scheyda et
al., in prep) with a limiting magnitude of $L = 16.2\,$mag and
contains about 3700 $L$-band sources. Those 201 sources for which we
have obtained $K$-band spectroscopy are marked and divided into four
groups: i) COES (blue), ii) COAS stars (red), iii) featureless
(yellow), and iv) others (green). Additionally, we have marked stars
with X-ray emission seen by Chandra by a vertical bar, and those with
infrared excess (IRE) by a horizontal bar. The number statistics is
summarised in Table\,1. The spatial coverage of our spectroscopy is
not homogeneous across the region, but focussed toward the cluster
centre and the southwestern interface between the \ion{H}{ii} region
and the molecular cloud.

\begin{table}
\caption{Number statistics of the four groups} \label{table:1}
\centering
\begin{tabular}{l@{\hspace{3.5mm}}c@{\hspace{3mm}}c@{\hspace{3mm}}c@{\hspace{3mm}}c}     
\hline\hline
Objects            & COAS      & COES   & Featureless & Others \\
\hline
Total              & 104       & 9        & 60          & 28        \\
IRE                & 34 (33\%) & 7 (78\%) & 32 (53\%)   & 10 (36\%) \\
X-ray            & 39 (38\%) & 4 (44\%) & 19 (32\%)   & 19 (68\%) \\
IRE and X-ray    & 19 (18\%) & 4 (44\%) & 10 (17\%)   & \ \ 6 (21\%) \\
IRE and/or X-ray & 54 (52\%) & 7 (78\%) & 41 (68\%)   & 23 (82\%) \\
\hline
\end{tabular}
\end{table}

\textbf{CO features --}
Altogether there are 9 (4\%) COES; 4 of them -- B331
(CEN\,92), B268 (CEN\,49), B275
(CEN\,24), and B337 (CEN\,93) -- were
previously known
(Hanson et al. 1997). We find 104 (52\%) COAS; \cite{hanson95} observed
4 of them -- B22, B120, B305 (CEN
102), and B324 (CEN\,33) -- and interpreted them as
field stars unrelated to M17. Two other COAS with IRE (B239,
B339) were suggested to be high-mass YSOs
(Hanson \& Conti 1995). Among these 113 CO sources, 109 do not show
any further spectral features. Furthermore, our sample contains 60
stars without stellar features, while the remaining 28
stars show photospheric absorption lines that allow their
classification as early-type stars.

\begin{figure}[h]
\resizebox{\hsize}{!}{\includegraphics*[width=\textwidth]{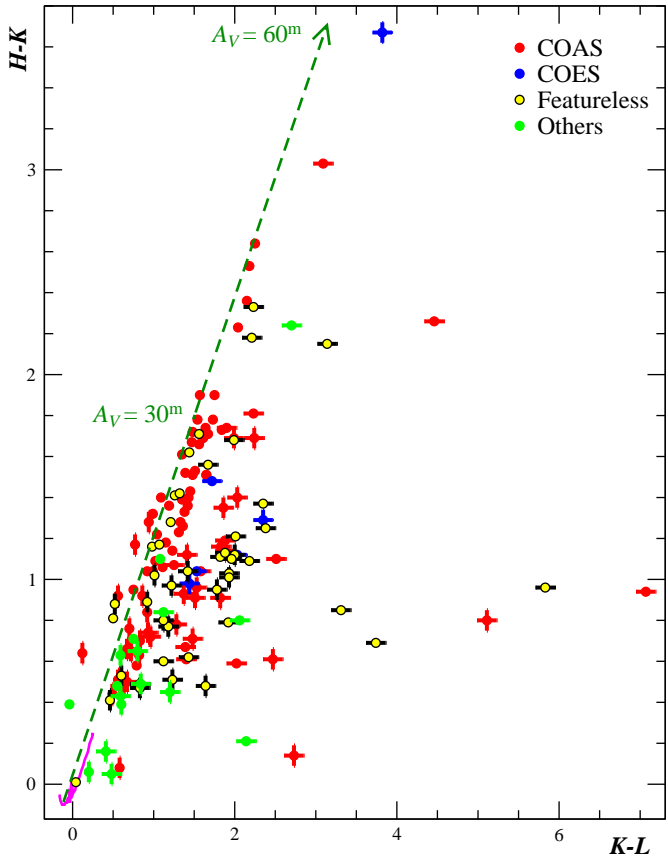}}
\caption{$HKL$ colour-colour diagram for 154 stars from our
spectroscopic sample. The notation is identical to that of
Fig.\,\ref{f1}. The solid curve (magenta) marks the locus of
unreddened main sequence stars; the dashed line (green) denotes the
reddening vector.
\label{f2}}
\end{figure}

\textbf{Infrared excess --}
Figure\,\ref{f2} shows the $HKL$ colour-colour diagram for 156 stars
from our spectroscopic sample; some of the brighter stars are
missing because they were saturated at one of the wavebands. Quite
a number of stars are located in the excess region below the
reddening path indicating thermal emission from dust in a
circumstellar disk and/or envelope. The infrared spectral energy
distribution for some of these objects has already been investigated
until $20\,\mu$m (Chini \& Wargau 1998; Nielbock et al. 2001)
and corroborates this
interpretation. The relative fractions of IRE among the four groups
shows a significant trend: IRE is present in 78\% of the COES, in
53\% of the featureless stars, in 36\% of the "others", and in 33\%
of the COAS.

\textbf{\emph{X}-ray emission --}
Within the cluster field of Fig.\,\ref{f1}, Chandra has detected 521
X-ray sources (Broos et al., in prep.). All of them are coincident
with infrared sources from our imaging survey. The Chandra sources
with CO features and with sufficient counts to support spectral fits
can be reproduced by a simple thermal plasma model, i.e., there is no
need for a flat power law, which is what one would expect if the
source were an X-ray binary or a background AGN. None of these sources
is anomalously soft and/or bright with a low absorbing column, so they
are unlikely to be foreground field stars; typical $A_V$ values range
between 5 and 25\,mag, consistent with our $JHKL$ photometry.  None of
these sources shows X-ray variability within the short (40 ksec)
observation. All of them are consistent with being members of the M17
complex.

Among our spectroscopic sample of 201 stars, there are 81
X-ray
sources. The distribution among the four groups is the following:
68\% of all stars in the group of ``others'' display X-ray
emission; according to their IR spectra these are mostly early--type
stars. The corresponding fractions in the other groups are: 44\%
(COES), 38\% (COAS), and 32\% (featureless).
The fractions of sources that simultaneously show X-ray emission
and IRE are 44\% (COES), 21\% (others), 17\% (featureless), and 18\%
(COAS).

\textbf{Sources of special interest --}
There are only four stars with the CO feature that have further
spectral lines at shorter wavelengths: B163, which was
reported to be featureless by \cite{hanson95}, is a COES that shows
Pa$\delta$ and Br$\gamma$ absorption in our spectrum. If this
discrepancy is not due to the low $S/N$ spectrum by \cite{hanson95} we
must assume that the source is variable. Additionally, B163 has both
X-ray emission and IRE. Another COES with IRE, CEN\,93
(Chini et al. 1980), also has Pa$\delta$ and Br$\gamma$ in absorption.
The COAS CEN\,57, with X-ray emission and IRE shows
Pa$\delta$ and Br$\gamma$ absorption, and \ion{He}{i} $\lambda$
21130\,\AA~in emission. CEN\,30, another COAS with X-ray
emission, shows Pa$\delta$, \ion{He}{i} $\lambda$ 10311\,\AA, and
Br$\gamma$ in absorption. In both cases, the infrared spectra are
consistent with early B-type stars.

%__________________________________________________________________

\section{Discussion}
\label{discussion}

Compared to previous CO band-head studies in star--forming regions,
the present survey is the most extensive one; in addition, it is the
only one that is fairly unbiased. \cite{casali96} suggested from
their sample of 44 YSOs in six regions that CO absorption is very
common in low-mass YSOs; from their Fig.\,2 we estimate the fraction
of COAS to be about 50\%. Likewise, \cite{casali96} argue that
Class\,II sources tend to show CO absorption, while Class\,I sources
are featureless. In the following, we will explore the nature of the
M17 sources in more detail.

\subsection{Cluster members or field stars?}

In accordance with previous studies we interpret the 9 COES as YSOs
within M17. Models for the thermal continuum emission from dusty
infalling envelopes around protostars indicate that the envelope
emission can exceed the stellar plus disk photospheric emission by
almost an order of magnitude (Calvet et al. 1997),
thus producing featureless spectra. The veiling of an envelope
weakens the CO absorption lines while a disk will only amplify the
CO feature. Therefore, the 60 featureless objects are very likely
Class\,I sources within M17. Those 54 COAS with X-ray
emission and/or IRE are also young cluster members, as demonstrated
above for the two objects CEN\,30 and 57. Only the nature of the
50 remaining COAS is a priori less clear because there is
no direct ``youth indicator'' for classification purposes. However,
there are several issues indicating that there might be
further YSOs in this group.

The first argument comes from the spatial distribution: most of the
COAS are located in the immediate vicinity of other young cluster
members and in regions with pronounced nebular emission. This makes
it likely that a considerable fraction of them is related to the M17
cluster. Secondly, the infrared colours of COAS without X-rays are
on average redder than for sources with X-ray emission (see
Fig.\,\ref{f2}); likewise, their apparent IR-brightness is fainter
by one or two magnitudes compared to the other stars in the sample.
Given that the scatter of intrinsic $HKL$ colours for all luminosity
classes is fairly small with respect to the observed colours in
Fig.\,\ref{f2}, one must assume that this large reddening is
primarily due to interstellar dust. Thus, all of them have
20 $<$ $A_V$\,$<$ 45\,mag, a fact that might explain the non-detections by
Chandra and that argues in favour of deeply embedded cluster members,
at least for some of them. Finally we want to note that the
groups of COES and featureless stars also contain 22\% and 32\%
of objects that neither show X-rays nor IRE; nevertheless, these
stars are most likely YSOs. Obviously, the absence of X-ray emission
or IRE is no conclusive argument against youth.

%__________________________________________________________________

\begin{figure}
\resizebox{\hsize}{!}{\includegraphics*[width=\textwidth]{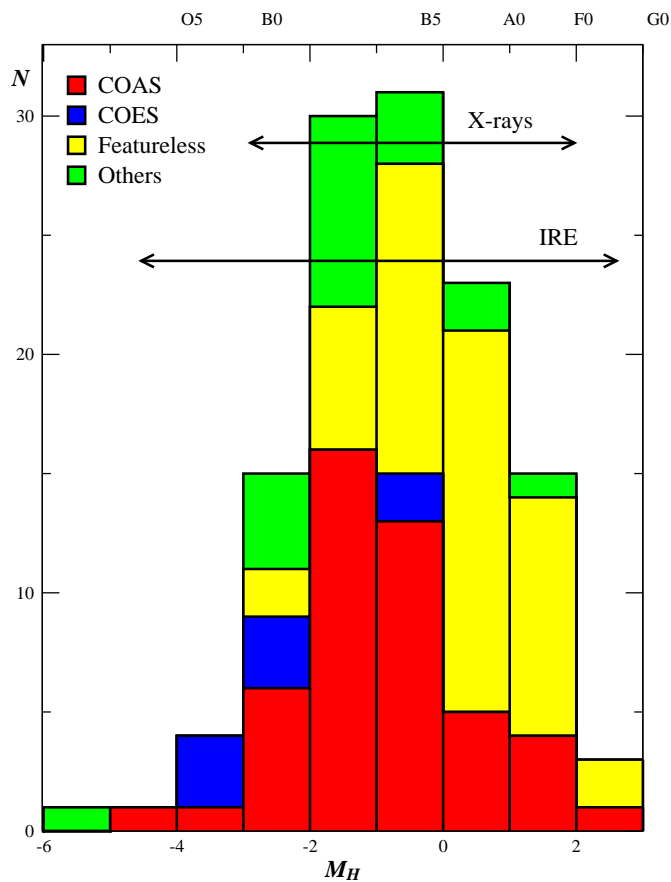}}
\caption{$H$-band luminosity function of probable cluster
members in M17. The colour code is identical to that of
Fig.\,\ref{f1}. The arrows denote the range where stars with X-ray
emission or IRE occur. \label{f3}}
\end{figure}

%__________________________________________________________________

\subsection{High or low-mass stars?}

The question concerning the mass of these YSOs is very important.
To estimate their luminosity we use the $J-H$ colours, where
contamination by dust emission is likely to be small. Dereddening is
achieved by adopting $(J-H)_0 \sim 0.16\,$mag, the mean intrinsic
colour of main sequence stars. As the range of intrinsic colours for
dwarf and giant stars is sufficiently small ($-0.1$ $<$ $(J-H)_0$ $<$
$0.3\,$mag for types O to K), the dereddening will not be affected much
by the stellar temperature. In this way, we obtain absolute magnitudes
of $-7.7 < M_H < 2.6\,$mag for the stars at the distance of
M17. This transforms  into intermediate to high-mass stars, mostly
earlier than A0, adopting main sequence luminosities. Some COAS that
can be independently dereddened on the basis of their X-ray spectra
corroborate this result: most of them appear to be early B-types; this
holds also for the COES.

Figure~\ref{f3} summarises the major statistical findings in the form of
an $H$-band luminosity function, adopting main sequence
luminosities. Those 50 COAS that do not show unique youth
indicators have been omitted from Fig.~\ref{f3}, although there might
be further YSOs among them as discussed above. The distribution
attains its maximum around B3; there are only 12\% A- and
2\% F-type stars. The remaining stars have absolute
$H$-magnitudes compatible with intermediate and high mass stars,
i.e., earlier than A0. Of course, at the faint end this
distribution is influenced by the limiting magnitude of our
spectroscopy. The faintest source for which a spectrum could be
obtained has an apparent $H$-band brightness of 17.6\,mag. Thus, it
follows that we can see stars of, e.g., type A3 only if their visual
extiction is below $A_V \sim 25$\,mag; types K6 can only be detected
with $A_V \le 5\,$mag.

Concerning the individual groups in Fig.~\ref{f3}, the featureless
objects are basically later than B3, while the COES are earlier than
B5. The COAS are distributed across the entire luminosity range.
The range $-5 < M_H < -2\,$mag contains 8 COAS that have X-ray
emission and/or IRE. This suggests the presence of YSOs with extremely
high luminosity.

The above results have assumed the stars to be on the main sequence,
although both CO band-heads and IRE indicate pre-main sequence
objects. While high-mass stars evolve mainly at constant luminosity
and therefore will not alter the overall distribution of spectral
types (or masses) in Fig.~\ref{f3}, low-mass pre-main sequence
objects change their luminosity by more than three orders of
magnitude and thus will simulate earlier types and higher masses. If
it is true that the $H$-band luminosity function is contaminated by
objects of lower masses, these objects must be in an extremely
early, i.e., protostellar evolutionary stage (due to their high
luminosity).

According to the model by \cite{calvet91} the "young" COAS,
i.e., those with X-ray emission or IRE, differ from the COES in the
sense that either the underlying YSO is cooler or the mass accretion
rate is higher. Corroborated by our early-type spectra for CEN\,30
and 57, we favour the latter interpretation and identify those COAS
as intermediate to high-mass protostars with mass accretion rates
above $10^{-5}\,$M$_\odot$\,yr$^{-1}$. Sources with featureless
spectra may be similar to other hot YSOs like the Herbig Ae/Be star
AB Aurigae (Hartmann et al. 1989)
and may form a population between the CO emission and absorption
sources.

\begin{acknowledgements}
We thank the Paranal Science Operations team for performing part of
the infrared observations in service mode. We also thank the Penn
State star formation group for providing results from their Chandra
data in advance of publication. Finally we thank our referee for
constructive suggestions. This work was partly funded by the
\emph{Nord\-rhein\--West\-f\"a\-li\-sche Aka\-de\-mie der Wis\-sen\-schaf\-ten} and by the
\emph{Deut\-sche For\-schungs\-ge\-mein\-schaft, DFG} through
SFB 591. N.V.
acknowledges support by FONDECYT grant 1061199 and by UCN grant
DGIP-10301180.
\end{acknowledgements}

%__________________________________________________________________

%\listofobjects

\Online
\begin{figure*}
\resizebox{\hsize}{!}{\includegraphics*[width=\textwidth]{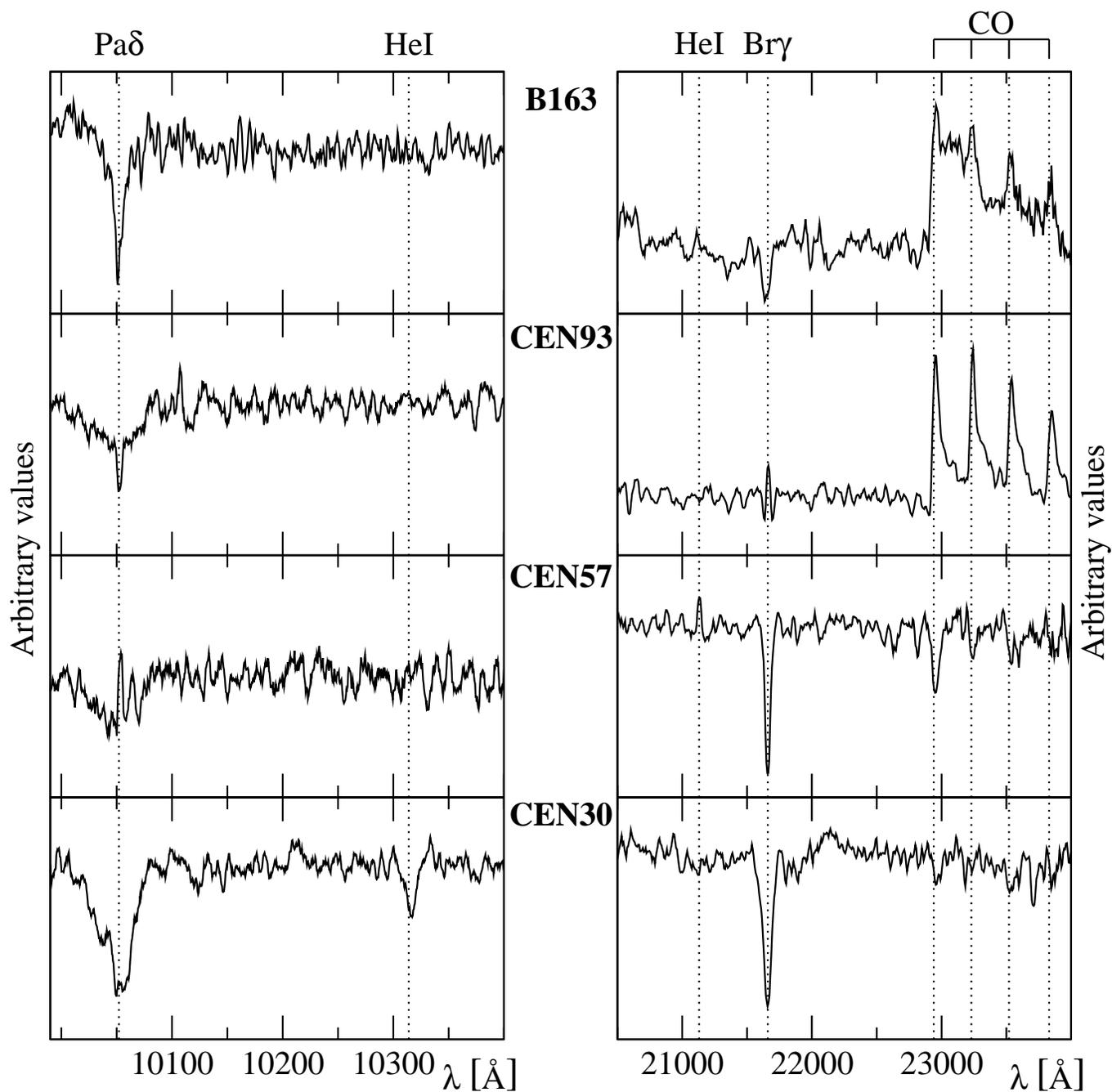}}
\caption{$J$- and $K$-band spectra of four objects, mentioned in the text
in the section "Sources of special interest". The positions of the CO
band-heads and several lines of H and He are marked.}
\label{of1}
\end{figure*}

\end{document}